\documentclass[%
reprint,
superscriptaddress,
 amsmath,amssymb,
 aps,
]{revtex4-1}

\usepackage{graphicx}
\usepackage{dcolumn}
\usepackage{bm}
\usepackage{hyperref}
\usepackage[mathlines]{lineno}

\begin{document}

\title{\textbf{Collective states of active particles with elastic dipolar interactions}}

\author{Subhaya Bose}

\affiliation{Department of Physics, University of California Merced, CA 95343, USA
}

\author{Patrick S. Noerr}%

 \affiliation{Department of Physics, University of California Merced, CA 95343, USA
}

\author{Ajay Gopinathan}
 \affiliation{Department of Physics, University of California Merced, CA 95343, USA
}%

\author{Arvind Gopinath}
\affiliation{%
 Department of Bioengineering, University of California Merced, CA 95343, USA
}

\author{Kinjal Dasbiswas}
\thanks{Correspondence: \\ Kinjal Dasbiswas \\ kdasbiswas@ucmerced.edu}
\affiliation{%
 Department of Physics, University of California Merced, CA 95343, USA
}%


\begin{abstract}
Many types of mammalian cells exert active contractile forces and mechanically deform their elastic substrate, to accomplish biological functions such as cell migration. These substrate deformations provide a mechanism by which cells can sense other cells, leading to long-range mechanical inter-cell interactions and possible self-organization.  Here, we treat cells as noisy motile particles that exert contractile dipolar stresses on elastic substrates as they move. By combining this minimal model for the motility of individual cells with a linear elastic model that accounts for substrate-mediated cell-cell interactions, we examine emergent collective states that result from the interplay of cell motility and long-range  elastic dipolar interactions. In particular, we show that particles  self-assemble into flexible, motile chains which can cluster to form diverse larger-scale compact structures with polar order. By computing key structural and dynamical metrics, we distinguish between the collective states at weak and strong elastic interactions, as well as at low and high motility. We also show how these states are affected by confinement – an important characteristic of the complex mechanical micro-environment inhabited by cells.  Our model predictions are generally applicable to active matter with dipolar interactions ranging from biological cells to synthetic colloids endowed with electric or magnetic dipole moments. 
\begin{description}
\item[Keywords]
\small{Mechanobiology, Cell Motility, Elastic dipole interactions, Active Brownian Particles,\\ Active Polymers, Self-organization, Brownian dynamics} 
\end{description}
\end{abstract}

\maketitle


\section{Introduction}

Active matter typically comprises autonomous agents,  biological or synthetic in origin, that harness internal energy sources to move \cite{Marchetti2013, Gompper2020}. These agents often undergo complex interactions with each other and their surrounding media that influence their collective behavior \cite{Bechinger2016}. Mammalian cells that move by crawling on elastic substrates such as tissue and constitute a canonical example of biological active matter in complex media, can cluster into persistently moving or rotating flocks \cite{Copenhagen2018}. These cells locomote by adhering to and exerting mechanical forces on their elastic extracellular substrate which they actively deform \cite{Ladoux2017, Alert2020}. The overall motility  is guided by cell-substrate substrate interactions as well as with other cells \cite{Angelini2010}. Cell-cell interactions can include mechanical interactions mediated mutual deformations of the surrounding substrate \cite{Reinhart2008, safran_13}. This is particularly the case for dilute cell cultures where cells are not in direct contact. On the other hand, in dense active matter systems such as in confluent epithelial cell monolayers,  direct cell-cell interactions including steric interactions can dominate  \cite{Henkes2020}. Mechanical interactions through a material medium are by their nature long-range and are expected to govern the collective states of active particles\cite{Chate2008}, and enrich the phenomena such as phase separation that arise from motility \cite{Cates2015, Digregorio2018}.

Mechanobiology studies of adherent cells cultured on elastic substrates \cite{balaban_01, Mandal2014}, suggest that substrate elasticity may provide a robust route to long lived and long ranged cell-cell interactions.
Indeed, cell culture experiments indicate that cells exert measurable forces on their neighbors, either through direct cell-cell contacts, or indirectly through mutual deformations of a compliant, extracellular substrate \cite{pelham_97, dembo_99}. The substrate-mediated elastic interactions between such cells has important implications for biological processes such as self-organization during blood vessel morphogenesis \cite{vanOers2014} and synchronization of beating cardiac muscle cells \cite{Tang2011, dasbiswas_15,Nitsan2016}. The overall motility of cells is modulated by cell-cell mechanical interactions, and therefore depends on substrate elasticity \cite{Reinhart-King2003}.

In general, active particles endowed with a dipole moment are expected to interact at long range with each other while also propelling themselves. Passive dipolar particles such as ferromagnetic colloids at equilibrium will align end-to-end into linear structures such as chains or rings \cite{deGennes1970, Nishiguchi2018}. 
At higher densities, the chains intersect to form gel-like network structures \cite{Ilg2011}. Topological defects in the chains such as junctions and rings are expected to affect the phases of passive dipolar fluids \cite{Tlusty2000, Rovigatti2011}.  When powered by chemical activity, dipolar colloidal systems exhibit self-assembly that depends on both the long-range, anisotropic interactions, as well as active motion, as revealed in recent experiments \cite{sakai2020active}. Such structures have also been studied in simulation in the context of active dipolar particles representing auto-phoretic colloids \cite{Kaiser2015, Liao2020}, as well as swimming microorganisms \cite{Guzman-Lastra2016} such as magnetotactic bacteria \cite{Telezki2020}. In related theoretical  studies, constrained or bundled chains of self-propelling colloidal  particles \cite{fatehiboroujeni2018nonlinear,  
sangani2020elastohydrodynamical, 
fily2020buckling, chelakkot2021synchronized} have also been 
shown to exhibit collective instabilities. Elasticity mediated interactions are seen to play critical roles, with the competition between mechanical interactions, steric interactions and activity determining the eventual dynamical behavior.

Here we build a minimal model of interacting elastic dipoles that is inspired by the mechanobiology of animal cells that actively deform their elastic substrate, while also exhibiting persistent motility. The starting point is the observation and deduction that contractile deformations of the underlying substrate originate from the elastic dipolar nature of stresses exerted by the cell on the substrate \cite{schwarz_02}.  We show that incorporation of these substrate-mediated interactions offers a robust way to the formation of compact, and relatively stable collective states. Our model combining active self-propulsion of the particles with their long-range dipolar interactions applies to a general class of  experimentally realizable systems, including synthetic colloids endowed with permanent or induced magnetic or electric dipole moments \cite{Yan2016}. By performing Brownian dynamics simulations on a collection of such dipolar active particles, we demonstrate the rich array of collective states that they can self-organize into. In particular, strong dipolar interactions promote end-to-end alignment of active particles, leading to self-assembled, motile chains. These chains can then further self-assemble into a hierarchy of larger-scale structures.

\section{Model}
\begin{figure*}[ht]  
		\centering
		\includegraphics[width=\textwidth]{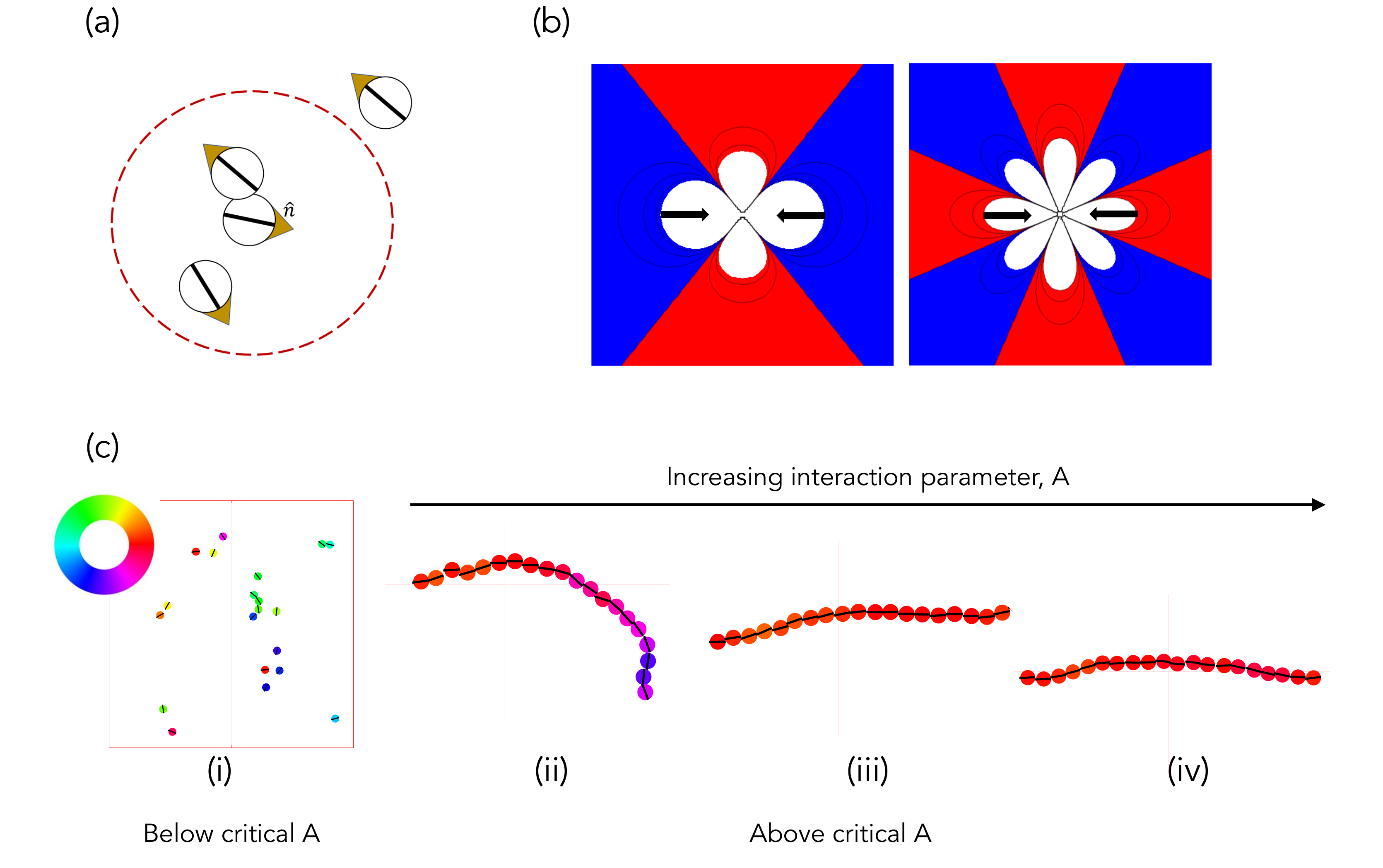}
		\caption{Overview of agent based simulations of active Brownian particles (ABPs) moving in the $x$-$y$ plane and interacting mechanically via elastic deformations induced by contractile,  active force dipoles. (a) Particles are modeled as circular ABPs  endowed with a dipole axis represented by the bold black line and an in-plane self-propulsion direction represented by the gold arrow. These move on a linearly elastic, thick, flat substrate, on which they exert contractile dipolar stresses. Substrate deformation due to one particle is sensed by neighbouring particles with these dipole-dipole elastic interactions confined to particles within a cutoff distance $r_{\mathrm{cut}}=7\sigma$ (shown as the dashed red circle). Particle overlap is penalized by a short range steric repulsion. (b) Representative spatial maps of the interaction potential 
		for a pair of contractile force dipoles in parallel (left) and perpendicular (right) configuration. The blue and red regions correspond to attractive and repulsive interactions, respectively. (c) (i) Simulation snapshot showing that weakly interacting particles do not stick to each other and move as independent entities. As the elastic dipolar interaction parameter $A$ increases, the particles self-assemble into long chains ((ii)-(iv), zoomed view shown). The flexibility of the chains and fluctuations in the mean curvature both decrease with increasing values of the interaction parameter.}
	\label{fig:Figure1} 
	\end{figure*}
Our model system consists of soft, repulsive, active Brownian particles (ABPs) \cite{Romanczuk2012, marchetti2016} in two dimensions (2D), that  interact at long range through elastic dipolar interactions and strongly repel when they overlap. We have previously studied a simple isotropic interaction model valid in the limit where the propulsion direction was decoupled from the magnitude of cell-cell interactions \cite{Bose2021}.
Here, we introduce and analyze a more general model that accounts for the anisotropy of cell interactions typically seen in biological settings. 

As shown in Fig.~\ref{fig:Figure1}a, the ABPs -- here termed particles -- are modeled as circular disks of diameter $\sigma$ that is endowed with a dipole moment and self-propulsion direction $\hat{\mathbf{n}}$. 
The orientation of $\hat{\mathbf{n}}$ is aligned with the dipole axis (shown as the bold black line). Particles move in this direction with self-propulsion speed $v_{0}$. Additionally, the motion of these particles are subject to random stochastic forces that mimic the effects of the thermal environment surrounding the particles. These random forces are the origin of diffusive effects in both orientation and spatial position of the ABPs.

Since we are motivated by adherent cells on elastic substrates whose contractile traction forces we model as elastic dipoles, a cutoff distance of $r_{\mathrm{cut}}=7\sigma$ (red dashed circle in Fig.~\ref{fig:Figure1}a) is imposed on the long range dipolar interactions. The choice of a cut-off length for interactions is consistent with experimental observations that cells can interact with one another via mechanical signalling at distances that are up to a few cell lengths away \cite{Reinhart2008, Tang2011}. In addition to the long-range interactions mediated by the elastic substrate, cells may also interact via short-range interactions. Here we introduce short-range steric repulsion using a mechanical  model using compressive springs that discourage overlap between neighbouring particles. Specifically, two particles in close-contact exert a repulsive elastic force on each other when the center-to-center distance is less than the rest length $\sigma$ of these springs.

 The model dynamics and interaction potential are detailed in the Methods Section~\ref{sec:Methods} in Eq.~\ref{potential_eq}-Eq.~\ref{eq:orientation}. Pairwise dipolar interactions are anisotropic and depend on both the distance between and relative orientations of the two particles with respect to their separation axis. Insights into the nature and scale of the elastic interaction potential between a pair of force dipoles, may be had by examining Fig.~\ref{fig:Figure1}b where we plot spatial maps of the potential fields for two canonical configurations. To plot these functions, we choose a  reference contractile force dipole that is fixed at the origin with axis along the $-x$ direction. This represents the contractile stresses exerted on the surface of a semi-infinite elastic substrate by an elongated, adherent cell. The  major axis  corresponds to both the dipole axis as well as direction of self-propulsion. A second test dipole, if present, interacts with the reference dipole according to its position and orientation, with the red (blue) regions in the potential maps in Fig.~ \ref{fig:Figure1}b representing repulsion (attraction).  While the map on the left corresponds to parallel alignment, that on the right maps the interaction potential for perpendicular alignment of the two dipoles. 
 The elastic material comprising the substrate is treated as a homogeneous material with shear and compression moduli both proportional to the Young's modulus, and a Poisson ratio $\nu$ that provides a measure of its compressibility \cite{Bischofs2003}. While our elastic model is valid for $ 0 < \nu < 1/2$, the figures plotted correspond to $\nu =0.1${\footnote{This choice ensures end-to-end alignment of dipoles and provides interactions seen not just in cells but also in other types of active matter that feature particles with magnetic or electric dipole moments. The interactions at $\nu > 0.3$ have a different symmetry and can result in more complex structures such as short rings without any electric or magnetic analogs \cite{Bischofs2006}.}}.

 The ensuing dipolar interactions, when strong enough relative to the stochastic noise, cause end-to-end chaining of the particles along their dipole axis. Examples of this chaining process are seen to occur in our simulations and representative snapshots are shown in figure \ref{fig:Figure1}c. As expected intuitively, increasing interaction gives rise to stronger alignment resulting in chains that are progressively less flexible. The effective elastic bending modulus of these chains that determines the fluctuations of the backbone contour of the chained ABP's is thus higher with increasing interaction strength.
 
 To illustrate the bulk behavior of interacting ABP's as well as the effect of confinement on emergent collective patterns, we simulate a few hundred of these particles in a box  confined in the $y-$ direction, and periodic in the $x-$ direction. This setup mimics a channel geometry typically used in cell motility experiments \cite{Marel2014-kd} and is used in other works on simulations of ABPs under confinement \cite{ ezhilan2015distribution, elgeti2015run, yan2015force}.  The important nondimensional control parameters in the model are the elastic dipole-dipole interaction strength $A$, the active self-propulsion velocity characterized by a P{\'e}clet number, $Pe$, and the packing fraction, $\phi$. The packing fraction used in simulations below is typically either $\phi =.08$ or $0.25$ corresponding to relatively dilute regimes, except in a narrow channel geometry where we go up to $\phi = .75$. Definitions and physical interpretations of these parameters are provided in the Methods Section~\ref{sec:Methods}.

\section{Results}

\subsection*{Characteristic states of active dipolar particles:  chains, polar bands, clusters and networks}

\begin{figure*}[ht]  
		\centering
		\includegraphics[width=\textwidth]{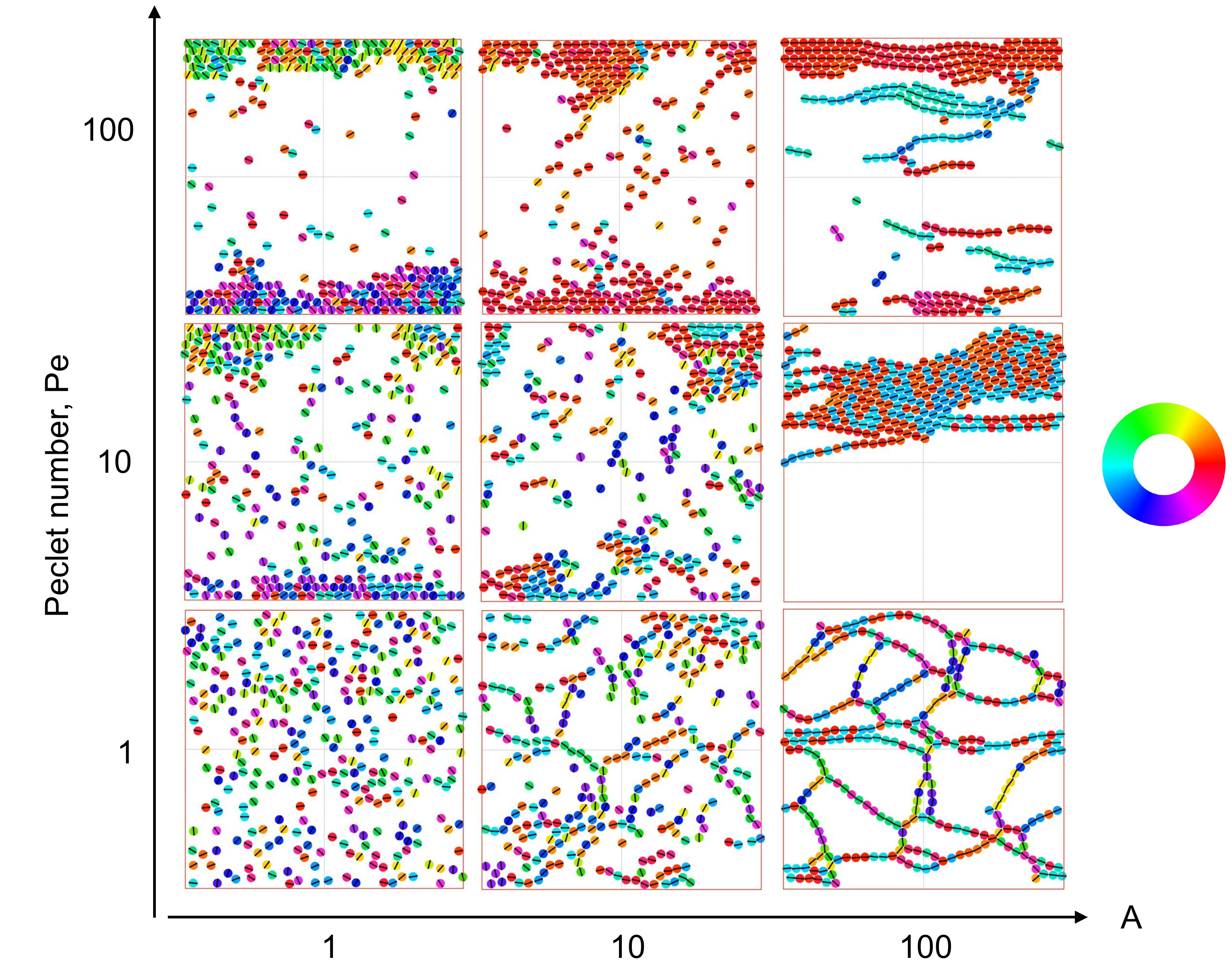}
		\caption{ Simulation snapshots of active particles with short range steric repulsions and long-range elastic dipole-dipole interactions as a function of effective elastic interaction $A = P^{2}/E\sigma^{3}k_{\mathrm{B}}T_{\mathrm{eff}}$ and P{\'e}clet number $Pe = \sigma v_{0}/D_{\mathrm{T}}$. Particles are confined in the $y$-direction, while they experience periodic boundary conditions in the $x$-direction. They are colored according to their self-propulsion direction $\mathbf{\hat{n}}$, and coded based on the color wheel. Motile particles at low effective elastic interaction collect into clusters at the boundaries. Strong elastic interactions promote network formation at low activity. Strong elastic interactions paired with high activity gives rise to active polymers and polar bands. The full movies corresponding to ABPs, networks, traveling cluster, and active polymers can be seen in Supplementary movies 1, 2, 3, and 4, respectively.}
		\label{fig:PhasePortrait} 
	\end{figure*}

\begin{figure*}[ht]  
	\centering
	\includegraphics[width=\textwidth]{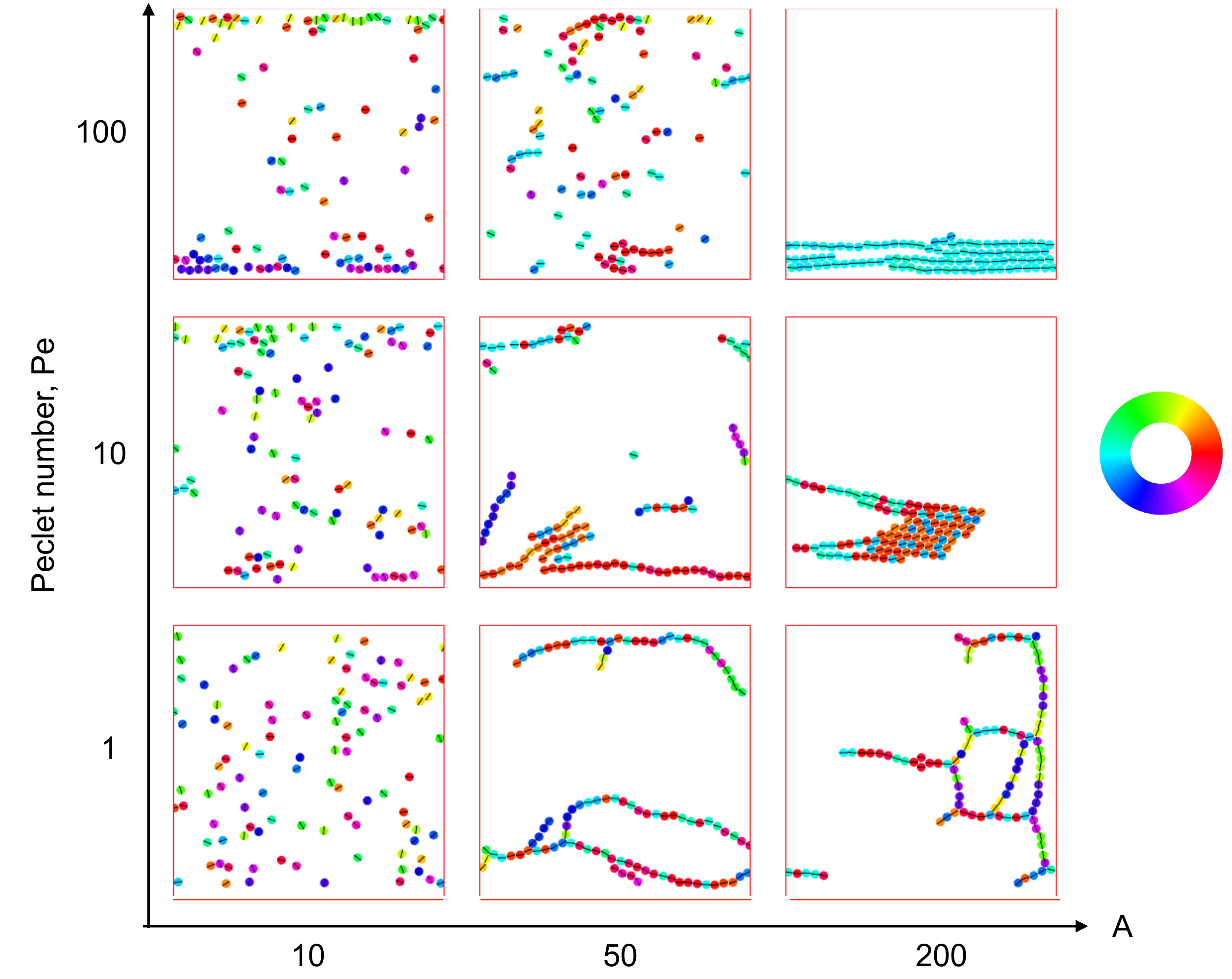}
		\caption{ Simulation snapshots of active particles at low packing fraction - The interaction parameter $A \equiv {P^{2}}/E\sigma^{3}k_{\mathrm{B}}T_{\mathrm{eff}}$ and P{\'e}clet number $Pe \equiv \sigma v_{0}/D_{\mathrm{T}}$ define the collective behavior of the particles. Particles are confined in the $y$-direction, while they experience periodic boundary conditions in the $x$-direction. They are colored based on the direction of $\mathbf{\hat{n}}$, as indicated by the color wheel.  At low interaction parameter $A = 10$, the particles remain isolated and diffuse (Supplementary Movie 6). At high $Pe$, more particles get collected at the confining boundary. At higher values of the interaction parameter, $A$, particles form chains. The typical length of the chains is seen to decrease with increasing $Pe$. At very high interaction parameter, $A = 200$, networks with multiple branches  form at low $Pe$, while chains aggregate into polar clusters at $Pe = 10$ (Supplementary Movie 7). Although the particles in the cluster are oriented in opposite directions, the cluster is stable and moves in the direction given by its overall polarity. Again at very high P{\'e}clet, $Pe = 100$, the particles in the chains are oriented in the same direction.}
		\label{fig:PhasePortrait_lowdensity} 
	\end{figure*}

We first explore the possible collective structures that result from the combination of active self-propulsion with dipolar attraction and alignment.  We explore the parameter space of activity (given by the P{\'e}clet number, $Pe$) and strength of dipolar interactions (given by the effective elastic interaction parameter, $A$) for two representative systems: one dilute and the other semi-dilute. We show representative snapshots of the steady states of the simulations by coloring the particles according their orientation. Collections of these snapshots as well as the color wheel corresponding to particle orientations are shown in Figs.~ \ref{fig:PhasePortrait}, where the packing fraction $\phi \approx .25$, and Fig.~\ref{fig:PhasePortrait_lowdensity}, where the packing fraction $\phi \approx .08$.

We see from figure \ref{fig:PhasePortrait} that at both low motility and weak elastic interactions($A=1$), particles do not form any ordered structures but are distributed uniformly in space, over the utilized simulation time. As motility is increased ($Pe \geq 10$), particles are seen to clump up at the boundary with their orientation vectors facing the wall at which they are localized. This is a familiar result of confined active Brownian particles (ABPs) wherein these tend to point towards the wall until their orientation is sufficiently randomized by the rotational diffusion\cite{Wagner2017}. As elastic interactions are dialed up such that the motions resulting from the dipolar interactions are much stronger than the stochastic diffusion of the system, structures characteristic of dipolar interactions emerge. In the case of low particle motility ($Pe = 1$), and high elastic interactions, we see a branched network form. In the case of intermediate motility ($Pe = 10$), networks are broken down into a single traveling cluster. In the former case, the particles comprising any given chain can either be oriented parallel ($0$) or anti-parallel ($\pi$) with respect to one another as the dipolar interaction is head-tail symmetric. In the latter case, networks form at short timescales and are compressed into one motile cluster at long timescales. This motile cluster contains numerous defects (shown by their different color) - particles oriented anti-parallel to the direction of cluster motion - caused by the earlier stage of network formation. Lastly, in the case of high particle motility ($Pe = 100$), particles assemble into traveling flexible chains. Much of our forthcoming analysis is focused on these highly ordered, yet highly dynamic, structures.

At low packing fraction (Fig. \ref{fig:PhasePortrait_lowdensity}), for $A = 10$ the elastic interaction between the particles is low and they diffuse around in the simulation space which is in contrast to what we see for higher packing fraction (Fig. \ref{fig:PhasePortrait}) where particles show alignment with weak attraction. Accumulation of the particles can be seen at the confining boundaries which is attributed to the activity of the particles. Upon increasing the elastic strength to $A = 50$, formation of chains has been observed.  At $Pe = 1$, long and branched chains of particles are formed. Increasing motility leads to a decrease in length of the chains and an increased polarity. At even higher elastic strength of $A=200$, long chains with multiple branches are seen for $Pe = 1$. At increased activity, the chains stick to each other and form an ordered cluster that moves coherently in the direction determined by the net polarity of the constituent particles. 

	
\subsection*{Pair correlations reveal spatial organization of active chains }

\begin{figure*}[ht]  
		\centering
		\includegraphics[width=\textwidth]{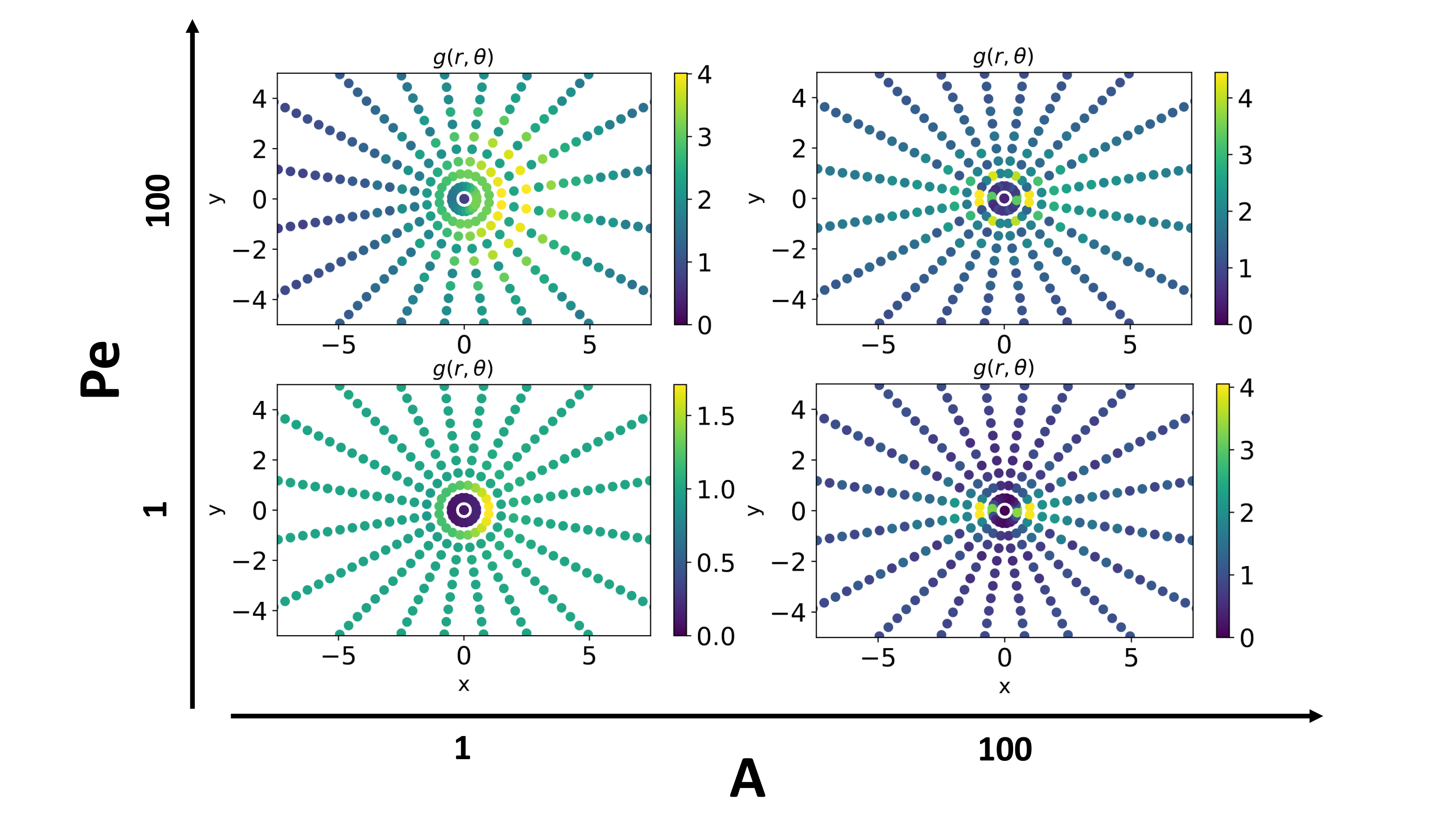}
		\caption{ Angular dependent pair correlation function is affected by both motility and elastic interactions. Strong elastic interactions promote pair correlation peaks at ($r,\theta$)$\:=\:$ ($\sigma$,$0$) $,$ ($\sigma$,$\pi$). At $Pe$, these are the only prominent peaks in the pair correlation function. Motile activity gives rise to secondary peaks at roughly ($r,\theta$)$\:=\:$($\sigma$,$\frac{\pi}{3}\:{\mathrm{mod}}\:\pi$)$,$ ($\sigma$,$\frac{2\pi}{3}\:{\mathrm{mod}}\:\pi$) as the preeminent structures are bundles of offset traveling chains. Weak elastic interactions broaden the pair correlation distribution. In this case, motility breaks head-tail symmetry and peaks can be seen at multiple integers of particle diameter at the head ($\theta = 0$ axis).}
		\label{fig:PairCorrelation} 
		
	\end{figure*}

To quantify the spatial distribution of particles around their neighbors, we calculate the pair correlation function, $g(r, \theta)$ and analyze the peaks in ($r,\theta$) space. Fig.~ \ref{fig:PairCorrelation} shows four such distance and angle dependent maps in the space of motility and elastic interaction. Elastic interactions localize the peaks of the pair correlation function. When motility is low, particles form branched networks and the primary configuration of particles is in straight chains. In this case, there exists two prominent peaks in the pair correlation function at ($\sigma$,$0$) and ($\sigma$,$\pi$). When both motility and elastic interactions are high, particles form into flexible traveling chains that have a tendency to join one another in a parallel fashion with an offset - a configuration that is energetically favorable to the elastic interaction and can be seen prominently in the simulation snapshot corresponding to $A=100$ and $Pe=100$ in Fig.~ \ref{fig:PhasePortrait}. In this case, the primary peaks still occur at ($\sigma$,$0$) and ($\sigma$,$\pi$), but secondary peaks are present at ($\sigma$,$\frac{\pi}{3}\:{\mathrm{mod}}\:\pi$) and ($\sigma$,$\frac{2\pi}{3}\:{\mathrm{mod}}\:\pi$), indicating the offset parallel band structure. Low elastic interactions constitute the more familiar case of collections of repulsive ABPs. In this regime, the head-tail symmetry characteristic of the elastic interactions is broken as particles are more likely to encounter other particles along their direction of propulsion \cite{Poncet2021}. There exists a single prominent peak at the head of the dipole that monotonically decreases on either side of the head axis. Increasing motility in the ABP system adds layers to the single peak function in integer multiples of particle size $\sigma$ as collision frequency increases.

\subsection*{Activity and elastic interactions promote orientational order}

\begin{figure*}[ht]  
		\centering
		\includegraphics[width=\textwidth]{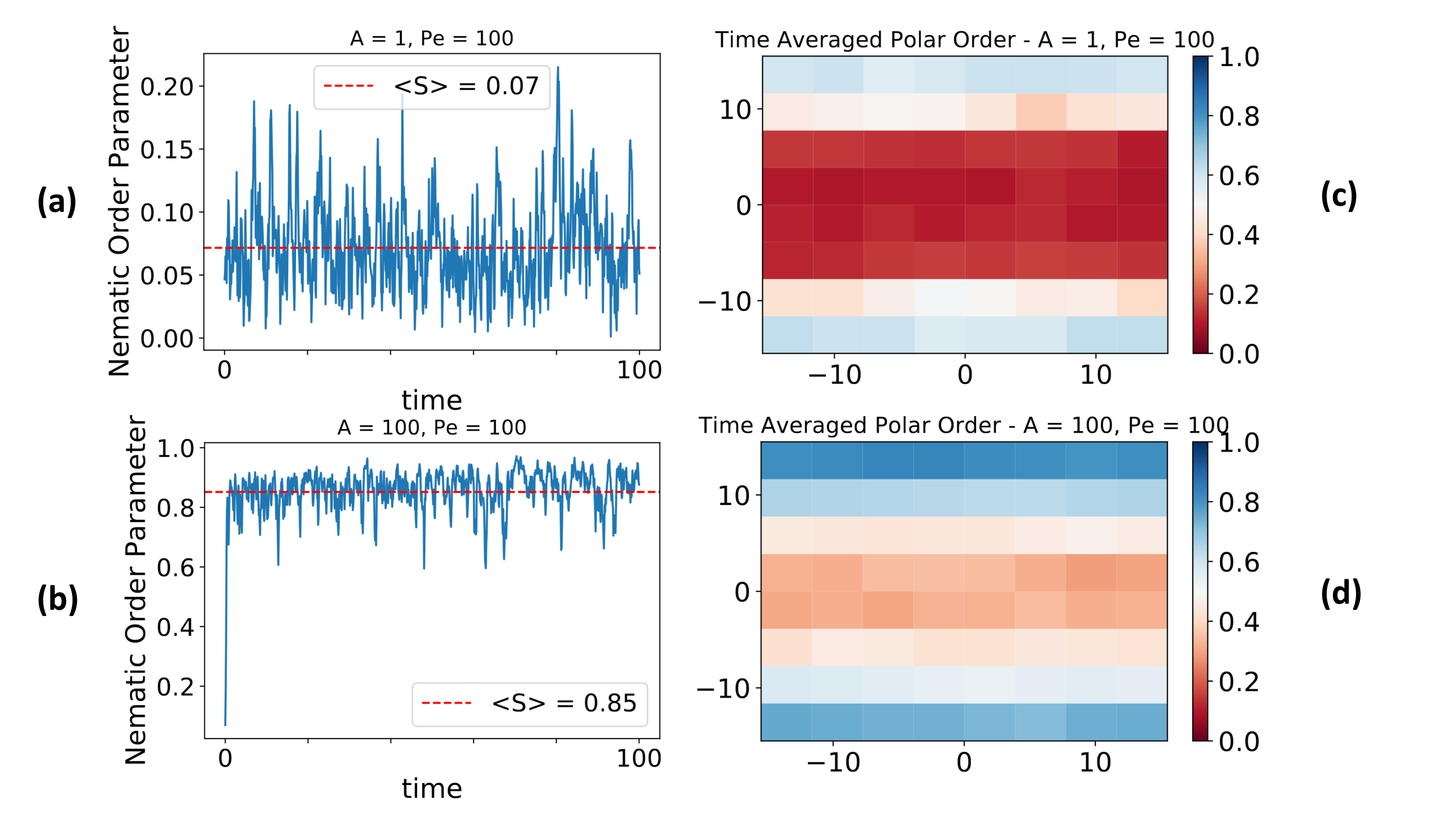}
		\caption{ Elastic interactions promote global nematic order and local polar order. (a) Global nematic order, measuring the overall alignment of the particles' dipole axes, vs. time for low effective elastic interaction and high activity. Average global nematic order is negligible for these parameters. (b) Global nematic order vs. time for high effective elastic interaction and high activity. The system quickly gains a persistent global nematic order parameter near unity because the chains align parallel to each other. (c) Spatial distribution of time averaged polar order, measuring the overall orientation of motility for the particles, for a characteristic run at low effective elastic interaction and high activity. Particles accumulate at the boundary and exhibit polar order along that boundary. This order rapidly decays away from the boundary  and there is virtually no polar order observed in the bulk. (d) Spatial distribution of time averaged polar order for a characteristic run at high effective elastic interaction and high activity. A polar order near unity is observed at the boundary and persists into the bulk where near the middle of the channel $|p| \approx 0.3$.}
		\label{fig:OrderParam} 
	\end{figure*}

At higher interaction strength, $A$, and higher motility, $Pe$, we see chains that move parallel to each other forming polar bands at high density (top right of Fig.~\ref{fig:PhasePortrait}). Since chains are elongated objects, a collection of them can give rise to orientational order, similar to active nematic and polar states that result from active, anisotropic particles \cite{Marchetti2013}. This type of order commonly seen in active matter comprising suspensions of cytoskeletal filaments and motors \cite{Winkler2020}. To quantify the orientational order in these cases and to distinguish from the individual ABPs under confinement, we measure the nematic and polar order for these states.  The magnitude of the nematic order parameter is defined as an average over the orientation of all particles, $S \equiv 2\langle \cos^{2}\theta \rangle-1$, where $\theta$ is the angle between a particle's orientation and the average director. In this case, the global alignment direction is parallel to the confining boundaries given by the $x-$ axis. The nematic order tells us how well the dipoles are aligned, without distinguishing between the head and tail and contains no information about the motility direction. To quantify the oriented motion, we  calculate the polar order, whose  magnitude is given by, $|p| \equiv \sqrt{ \langle n_{x}\rangle^{2}+\langle n_{y} \rangle^{2}}$, where $n_{x}$ and $n_{y}$ are the $x$ and $y$ components of the orientation vector, $\hat{\mathbf{n}}$, respectively. This quantity is higher if the particles are oriented in the same direction, in addition to being aligned. While nematic alignment is encouraged by the passive dipolar interactions, active motility induces polar order.

 Fig.~\ref{fig:OrderParam}(a,c) shows the global nematic order in time and Fig.~\ref{fig:OrderParam}(b,d) shows the time averaged spatial map of the polar order parameter for both ABPs and traveling flexible chains. In the ABP system, the global nematic order is small due to the tendency of particles at the walls to be oriented orthogonal to the wall and those in the bulk to be oriented  parallel to the wall, as well as the presence of orientational fluctuations from rotational diffusion. Traveling flexible chains of dipolar particles exhibit a global nematic order close to unity as all particles in this system tend to point along a director parallel to the confined boundary. Spatially resolving the average of the magnitude of the polar order parameter gives us a picture of particle alignment at a smaller length scale. ABPs exhibit polar alignment at the boundary. This alignment quickly diminishes and no polar order is seen in the bulk. Traveling chains form bands at the boundary such that $p>.7$ up to $6\sigma$ away from the wall. The polar order of these flexible chains drops off far less drastically in the bulk than the ABP system.

\subsection*{Transport properties of active chains are distinct from single particles}

\begin{figure*}[ht]  
		\centering
		\includegraphics[width=\textwidth]{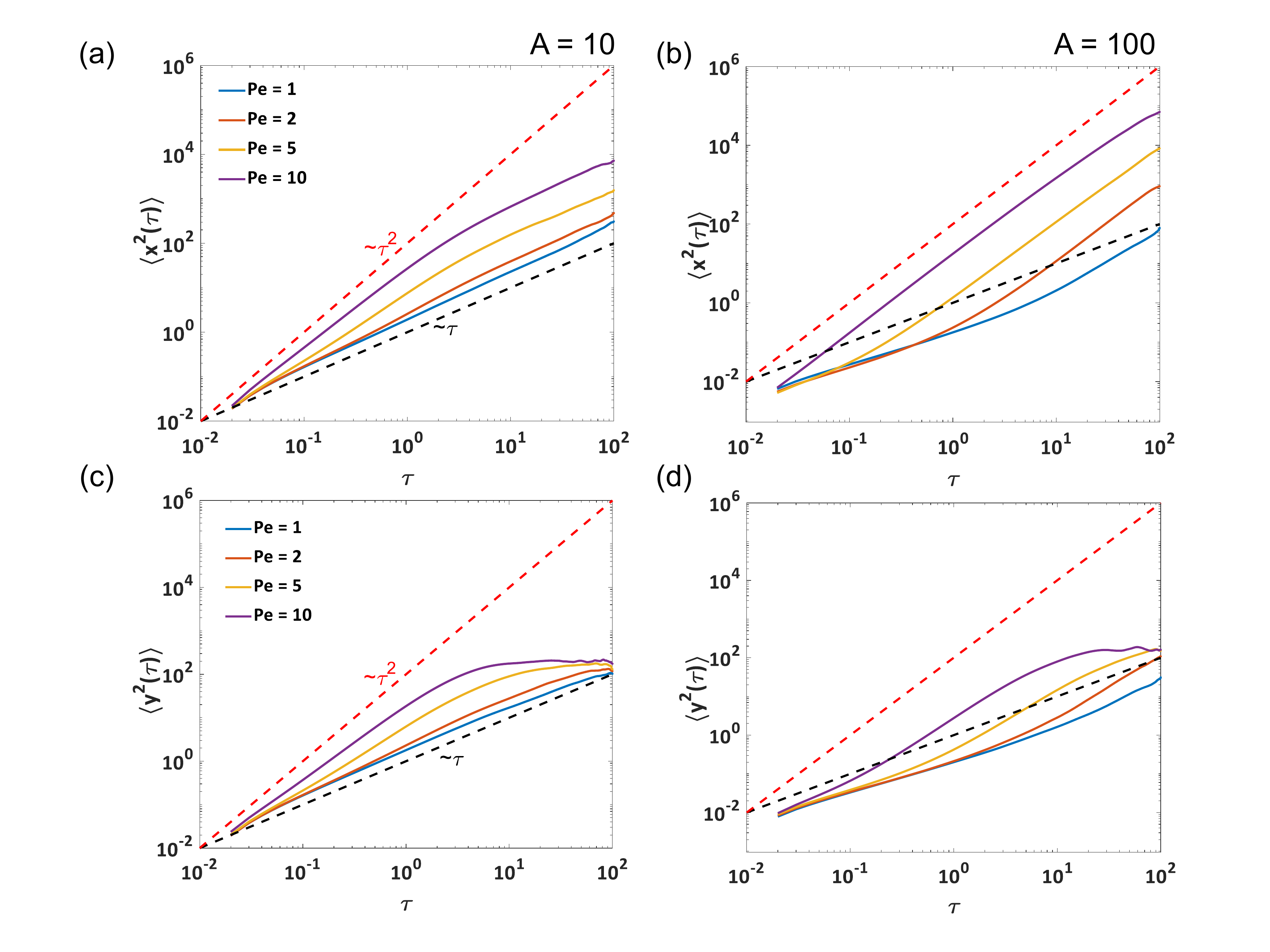}
		\caption{Mean-Squared Displacement or MSD vs. time interval, for 100 particles in a square simulation box of 30$\sigma$. Due to confinement of particles in y-direction, MSD is plotted separately for $x$ and $y$ components of displacement. (a), (b) MSD along unconfined direction: for $A = 10$, particles are super-diffusive at short time scale and diffusive at longer time scale, where the crossover time scale is determined by the P{\'e}clet number ($Pe$) of the particles. At $A = 100$, particles align themselves to form chains or clusters. At low $Pe$, the particles show sub-diffusive behavior at shorter times and ballistic behavior at longer times. At higher $Pe$, the ballistic behavior of particles is observed at all time scales. (c), (d) MSD along confined direction: particles reach the confining boundary at shorter times for high $Pe$ number, and also at low elastic interactions $A$. At higher $A$, particles chain up and move predominantly parallel to the confining boundary.}
		\label{fig:MSD} 
	\end{figure*}

The mean-squared displacement or MSD is a typical metric that quantifies how motile entities cover space in time. In Fig.~\ref {fig:MSD}, we report the MSD for simulations with a packing fraction $\phi \approx 0.08$ in a square box of size $30\sigma$, corresponding to the structures shown in Fig.~\ref{fig:PhasePortrait_lowdensity}. Given the confinement along one direction, we calculate the MSD separately for the confined ($y$-) and unconfined ($x$-) directions. The unconfined MSD, $\langle x^2 \rangle$, for particles with low elastic interaction \textit{e.g.}, at $A = 10$ , shows similar trends to individual active Brownian particles \cite{Volpe2014}. At short time intervals, individual ABPs propel persistently in the direction of their orientation, leading to ballistic behavior. In Fig.~\ref{fig:MSD}a, we see such behavior at very short time scales which gave way to super-diffusive behavior at intermediate time scales, where particles are slowed down by collisions with other particles. At sufficiently long time scales, the particles are diffusive as the rotational diffusion randomizes their orientation. Increasing P{\'e}clet number increases the timescale for super-diffusive behavior as the persistence time is longer. 

We see qualitatively different regimes in the MSD for particles with stronger interaction in Fig.~\ref{fig:MSD}b. At interaction strength $A \geq 100$, which leads to formation of long, stable chains, we observe larger-scale structures such as branches, clusters and networks in the simulation snapshots shown in Fig.~\ref{fig:PhasePortrait_lowdensity}. In this case, the particles show sub-diffusive behavior at shorter time scales when they are still moving individually in an uncorrelated manner and beginning to form these structures. On the other hand, at longer time scales, they cluster into larger scale structures that  move coherently in a specific direction like polar flocks, giving rise to a ballistic behavior. The crossover from subdiffusive to nearly ballistic behavior occurs earlier for higher P{\'e}clet numbers. At higher particle motility, we obtain ballistic behavior for all time scales. The resulting behavior is thus qualitatively different from single ABP behavior, which show a crossover from persistent to diffusive motion at timescales longer than the persistence time ($\sim Pe$). Here, on the other hand, the long time behavior is dictated by large-scale, polar structures that self-assemble irreversibly and move persistently at long times.

The MSD in the confined direction, $\langle y^2 \rangle$, plateaus off at long times, both for the individual ABPs (Fig.~\ref{fig:MSD}c) and the larger scale structures (Fig.~\ref{fig:MSD}d). The time scale to reach a plateau in the MSD corresponds to the time it takes an entity to reach the confining walls from the bulk of the simulation box. Thus,  $\langle y^2 \rangle$ reaches a plateau at a shorter time scale for highly motile particles, as compared to the less motile ones. Due to the confining wall in the $y$-direction and strong alignment with neighboring particles at $A = 100$, the particles line up into chains that orient and move parallel to the confining walls, and not as much in the $y$-direction. Thus, $\langle y^2 \rangle$ for $A = 100$ reaches the plateau later than for the $A =10$ case, for corresponding values of $Pe$.

\begin{figure*}[ht]  
		\centering
		\includegraphics[width=\textwidth]{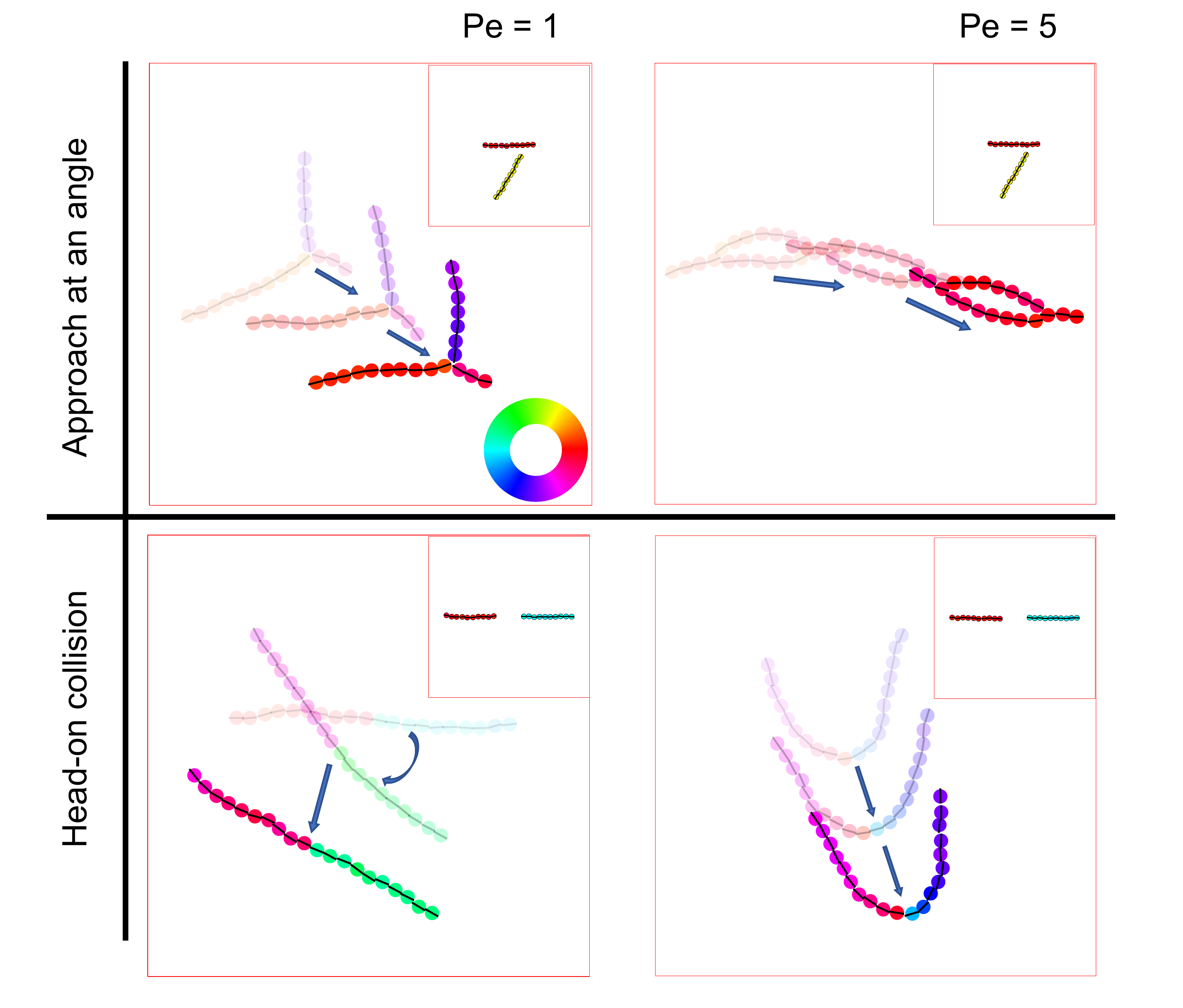}
		\caption{Interaction of two motile chains (Supplementary Movie (Supplementary Movie 8). Two straight chains of $10$ particles each are initialized to approach each other at an angle of $\frac{\pi}{3}$ and also $\pi$ (``head-on'') at $Pe = 1$ and $5$. At $Pe = 1$, a `Y' junction forms for an approach angle of $\frac{\pi}{3}$ whereas at $Pe =5 $, an `eye' ( two junctions) occurs. Upon head-on collision, a longer fluctuating chain with negligible net motility results at $Pe = 1$, and a propelling, buckled shape is observed  at $Pe = 5$. Insets at the top corners represent the approach of the chains. Color represents angle of orientation of particles. The arrows indicate progression in time and suggest that the configurations are both stable and motile.}
		\label{fig:ChainInt} 
	\end{figure*}

\subsection*{Collisions of active chains reveal stable, mobile structures }

We observe from simulations at low packing fraction (Fig.~ \ref{fig:PhasePortrait_lowdensity}) that once particles self-assemble into chains, these can intersect to form junctions and get organized into larger-scale polar structures. We now explore in more detail the inter-chain interactions responsible for this self-organization. To do this, particles were initialized in an ordered chain and oriented in the same direction. Two such chains were oriented initially at different angles to control their approach direction, as shown in the insets in Fig.~\ref{fig:ChainInt}. 

At $A = 200$ the junctions formed by chains depended on the P{\'e}clet number and the angle and position of approach. The `Y’ junction was the most observed for all P{\'e}clet number, which is formed from when the second chain attaches itself at the middle of the first chain (Figure \ref{fig:ChainInt}, top left). An `eye’ (Figure \ref{fig:ChainInt}, top right) is formed from two closely spaced `Y’s, which is observed for higher P{\'e}clet number, $Pe = 5$ and $10$ and when the chains are oriented in the same direction. Again, at low particle motility $Pe = 1$, the chains upon colliding head on form a longer and more rigid chain (Figure \ref{fig:ChainInt}, bottom left). On the other hand, at $Pe = 5$ chains show buckling upon undergoing head on collision forming a propelling `necklace’ (Figure \ref{fig:ChainInt}, bottom right). At even higher P{\'e}clet number, the force between the particles is overpowered causing particles to detach from a chain and thereby creating defects. All these cases have been observed for $A = 200$. These junctions are also observed at lower elastic strength $A = 50$ and $100$, but were unstable giving rise to many defects. Chains may interact with each other in a head-tail fashion which results in a stable longer chain. Chains with multiple defects have also been observed to form these `Y' and `eye' structures at $A = 200$ and $Pe = 1$ (Figure \ref{fig:PhasePortrait_lowdensity}).

\subsection*{Stronger confinement in narrow channels reveals polar clustering dynamics}

\begin{figure*}[ht]  
		\centering
		\includegraphics[width=\textwidth]{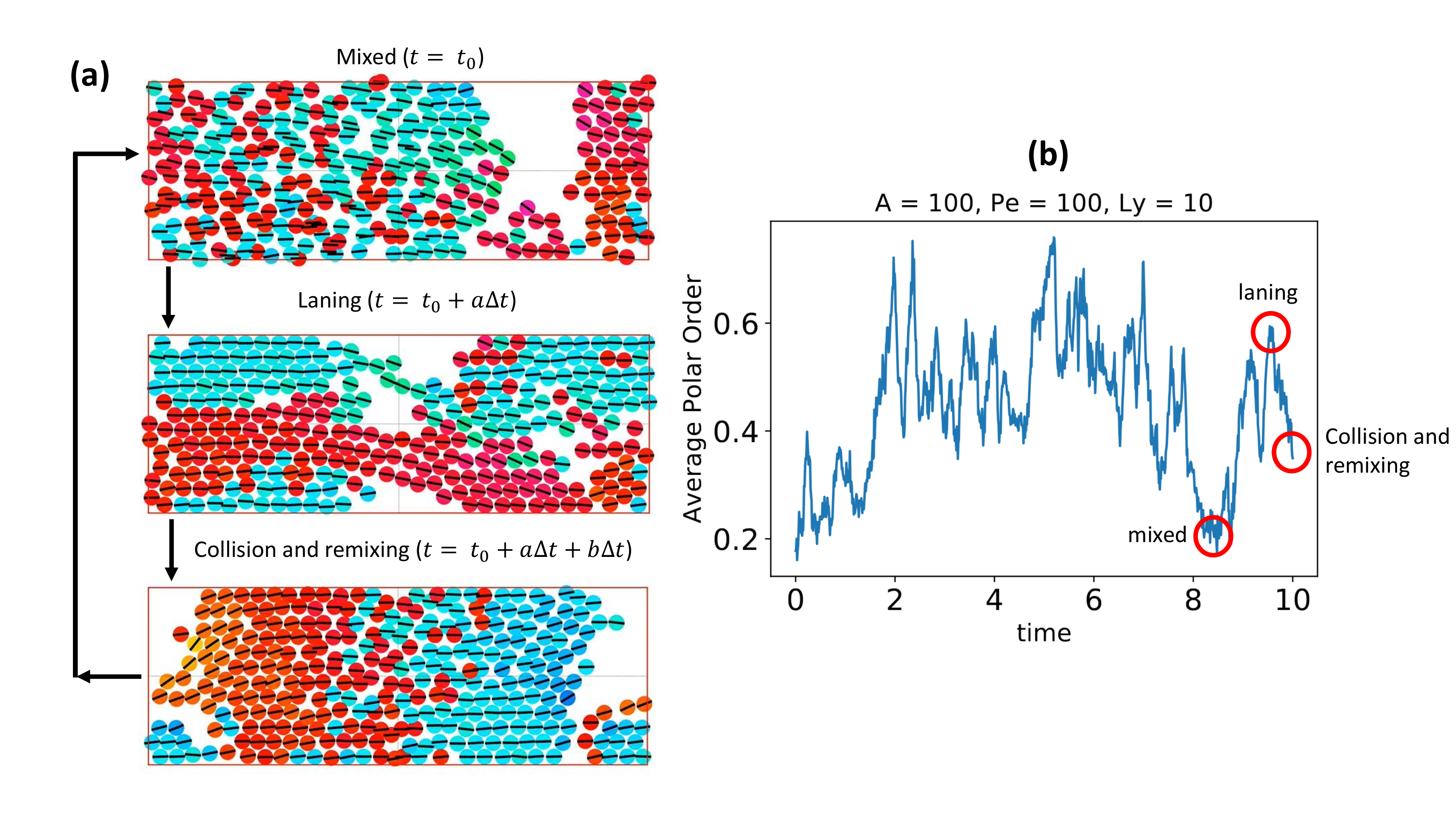}
		\caption{ Traveling chains in a narrow channel exhibit cycles of mixing, laning, and collision and remixing. (a) Snapshots of a  simulation where channel width has been decreased by a factor of $3$. Dynamics of the system are encompassed by three cyclic states: A mixed state shown at an arbitrary initial time $t_{0}$, a phase separated laning state shown a short time after $t_{0}$, and a collision and remixing state shown a short time after the laning state. (b) Average polar order versus time is shown to elucidate the three states described in (a). The polar order for a mixed - laning - collision and remixing cycle are shown in the red circles. When the system is well mixed, the average polar order is small ($p \approx .2$). When the particles separate into lanes, the polar order increases rapidly ($p \approx .6$). When the lanes then collide and begin remixing, the remnants of the bulk of the lanes provide polar order while mixed particles and the interface between lanes decreases polar order ($p \approx .4$). The full movie corresponding to the narrow channel can be seen in supplementary movie 5.}
		\label{fig:ChannelWidth} 
	\end{figure*}

In our system of traveling flexible chains comprised of strongly interacting and highly motile dipoles ($A=100, Pe=100$), bands that form along the boundary are relatively stable compared to those that form in the bulk. These latter are subject to more frequent collisions with other traveling chains. In order to gain understanding of these chain collision dynamics, we confine our particles into a channel of width $\frac{L}{3}$, where $L$ is the box size of our original simulation space, in order to induce more frequent and global chain-chain collisions. In this system we find a cyclic tripartite state dynamic. As shown in figure \ref{fig:ChannelWidth}a, at some point, the particles with orientations $+ x$ become well mixed with particles with orientations $-x$. The particles will then be separate into lanes according to their polarity so that they can move unimpeded. These lanes will then collide which initializes another well mixed system and the cycle repeats. This can be seen quantitatively by tracking the average magnitude of the polar order parameter in time shown in figure \ref{fig:ChannelWidth}b. The well mixed system has an average polar order parameter of $p \approx .2$. The system then phase separates into lanes with average polar order parameter $\approx .6$. The $+x$ and $-x$ lanes collide and the resultant combination has an average polar order parameter $\approx .4$. This time dependent formation and disbanding of polar structures is consistent with bead spring simulations of semiflexible filaments in the high activity regime \cite{Vliegenthart2020}. 

\section {Methods} \label{sec:Methods}

Here we present the equations governing the motion of the active motile particles discussed earlier and their interaction via the elastic substrate on which they move.
In our model, we treat the particles as circular active Brownian particles (ABPs) that interact with other particles via long-range substrate modulated interactions and direct short-range particle-particle steric contact interactions. 
Long range interactions arise as each ABP contractile stress dipoles ${\bf{P}}$ on the flat, semi-infinite, linearly elastic, isotropic substrate thereby inducing strain fields which affects the motility of nearby particles. For simplicity, we assume that the dipole axis is coincident with the direction of motion of the particle. For instance in an elongated cell, the force dipole axis coincides with the orientational axis of the cell, that is also the direction of self-propulsion.

In the derivation that follows we use Einstein summation convention over the Latin indices, while Greek indices are used to label the particles. Consider a particle $\alpha$ interacting with and deforming the substrate. The work done by the associated dipole, $\mathbf{P}_{\alpha}$ in deforming the substrate in the presence of the strain created by second dipole $\mathbf{P}_{\beta}$ (generated by a second particle $\beta$)  is given by 
\[
W_{\alpha\beta} = P^{\beta}_{ij} \partial_j \partial_l G^{\alpha \beta}_{ik}({\bf r}_{\alpha \beta}) P^{\alpha}_{kl},
\]
where ${\bf r}_{\alpha \beta} = {\bf r}_{\beta}-{\bf r}_{\alpha}$ is the separation vector connecting the centers of particles $\alpha$ and $\beta$ (c.f \cite{schwarz_02, Bischofs2003}).  The elastic half space Green's function that captures the displacement fields and thus the associated strain function in the linearly elastic medium at the location of one particle (dipole) caused by the application of a point force at the location of the other is given by \cite{landau_lifshitz_elasticity},
\begin{equation}
G^{\alpha \beta}_{ik}({\bf r}_{\alpha \beta}) = \frac{1+\nu}{\pi E} \bigg[(1-\nu) \frac{\delta_{ik}}{r_{\alpha \beta}} + \nu \frac{r_{\alpha \beta,i} r_{\alpha \beta, k}}{r_{\alpha \beta}^3} \bigg],
\label{Boussinesq:Greensfunction}
\end{equation}
where $E$ is the stiffness (Young's modulus) and $\nu$ is Poisson's ratio of the substrate. Given the linearity of the problem, superposition of strain fields each of which is obtained by using the Green's function (1) appropriately provides the net displacement at a test position due to particles around it. 

Two particles in our model interact via a combination of pairwise long-range and short-range interactions. The long range interaction forces originate from the elastic dipole-dipole interaction and can be derived from the elastic potential by using the Green's function and the equation for the work $W_{\alpha \beta}$. The short-range interactions are steric in nature and prevent ABP's from overlapping. 
This functionality is achieved in the framework of our model by the use of 
linear spring that only resist compression. Taken together, the total interaction potential between particles $\alpha$ and $\beta$ can be written as,
\begin{eqnarray}
    W_{\alpha \beta} &=& 
        {1 \over 2}k {(\sigma - r_{\alpha \beta})^2}, \:\:\: {\mathrm{when}} \:\:\: 0\leq r_{\alpha \beta}< \sigma \:\:\: \:\:\:  \nonumber \\   
        &=&
        {P^{2} \over E} {f(\nu,\theta_{\alpha},\theta_{\beta}) \over r_{\alpha \beta}^{3}}, \:\: \:\:\: {\mathrm{when}} \:\:\: \sigma\leq r_{\alpha \beta}< r_{\mathrm{cut}} \:\:\: \:\:\:  \nonumber \\   
        &=&0,\:\:\: {\mathrm{when}} \:\:\:  r_{\alpha \beta}\geq r_{\mathrm{cut}}.
    \label{potential_eq}
\end{eqnarray}
where $k$ is the spring constant of the linear (repulsive) spring preventing overlap, $\sigma$ is the particle diameter (kept constant in our simulations),  and $r_{\mathrm{cut}}$ is a cutoff distance beyond which the dipolar interactions are neglected. The magnitude of each force dipole is taken to be the same value denoted by $P$. The dependence of the pairwise dipolar interactions on the orientations of the two dipoles with respect to their separation vectors, and on the Poisson ratio of the medium, $\nu$, is expressed compactly in the expression \cite{Bischofs2003},
\begin{align}
f(\nu,\theta_{\alpha},\theta_{\beta}) = & \frac{\nu(\nu+1)}{2\pi} \Big(3(\cos^{2}\theta_{\alpha}+\cos^{2}\theta_{\beta}\\ \nonumber 
&-5\cos^{2}\theta_{\alpha}\cos^{2}\theta_{\beta}-\tfrac{1}{3}) \\ \nonumber
&-(2-\nu^{-1})\cos^{2}(\theta_{\alpha}-\theta_{\beta})\\ \nonumber 
&-3(\nu^{-1}-4)\cos\theta_{\alpha}\cos\theta_{\beta}\cos(\theta_{\alpha}-\theta_{\beta})\Big).
\end{align}
where $\cos \theta_{\alpha} = \bm{\hat{n}}_{\alpha}\cdot{\vec{\bm{r}}_{\alpha \beta}}$ and $\cos \theta_{\beta} = \bm{\hat{n}}_{\beta}\cdot{\vec{\bm{r}}_{\alpha \beta}}$ are the orientations of particles, $\alpha$ and $\beta$, with respect to their separation vector, respectively.

Motivated by natural and synthetic systems to which our model is applicable,  we assuming that the particles are in an over-damped viscous environment, and  inertia of the ABP's can be ignored. We can then write the equations of motion governing the translation and rotation, respectively, of particle $\alpha$ as,
\begin{equation}
    \frac{d\mathbf{r}_{\alpha}} {dt} = v_{0}\hat{\mathbf{n}}_{\alpha} -\mu_{T}\sum_{\beta} \frac{\partial W_{\alpha \beta}}{\partial {\bf r}_{\alpha}} + \sqrt{2D_{\mathrm{T}}}\:\: \mathbf{\eta}_{\alpha,\mathrm{T}}(t)
\label{eq:translation}    
\end{equation}
and
\begin{equation}
    \frac{d\hat{\mathbf{n}}_{\alpha}} {dt} =  -\mu_{R} \sum_{\beta}\hat{\mathbf{n}}_{\alpha}\times\frac{\partial W_{\alpha \beta}}{\partial \hat{\mathbf{n}}_{\alpha}} + \sqrt{2D_{\mathrm{R}}}\:\: \bm{\eta}_{\alpha,\mathrm{R}}(t),
\label{eq:orientation}    
\end{equation}
where $\mathbf{r}_{\alpha}$ and $\hat{\mathbf{n}}_{\alpha}$ are the position and orientation of particle $\alpha$, respectively. In the equations above $D_{\mathrm{T}}$ and $D_{\mathrm{R}}$ are the translational and rotational diffusivity quantifying the random motion of a single particle, respectively. The viscous environment results in the translational and rotational mobilities, $\mu_{\mathrm{T}}$ and $\mu_{\mathrm{R}}$  respectively. Random white noise terms $\bm{\eta_{\mathrm{T}}}$ and $\bm{\eta_{\mathrm{R}}}$ have components that satisfy $\langle \eta_{i,\mathrm{T}}(t) \eta_{j,\mathrm{T}}(t') \rangle = \delta(t-t') \delta_{ij}$ and $\langle \eta_{i,\mathrm{R}}(t) \eta_{j,\mathrm{R}}(t') \rangle = \delta(t-t') \delta_{ij}$. Since the fluctuation dissipation theorem is not necessarily satisfied for a nonequilibrium system, the translational and rotational diffusivity are independent of each other. However, to reduce the number of free parameters and in the interests of simplicity, we assume that  $D_\mathrm{T}=\sigma^{2}D_\mathrm{R}$ and $\mu_\mathrm{T}=\sigma^{2}\mu_\mathrm{R}$.  This allows the definition of an effective temperature, $k_{B} T_{\mathrm{eff}} =  D_{\mathrm{T}}/\mu_{T}$.  Finbally we emphasize that each particle is endowed with the same dipole strength, $P$, and self-propulsion velocity, $v_{0}$, both of which are constant. 

We now choose the cell diameter $\sigma$, the diffusion time, $\sigma^{2}/D_{\mathrm{T}}$, and the effective thermal energy that quantifies the strength of stochastic fluctuations, $D_{\mathrm{T}}/\mu_{T}$, as physically relevant length, time, and energy scales in our model. Solutions to the scaled dynamical model are then 
dependent on three non-dimensional numbers,
\begin{equation}
Pe = \frac{v_{0}\sigma}{D_\mathrm{T}}, \, A = \frac{\mu_{T}P^{2}}{E\sigma^{3}D_{\mathrm{T}}}, \, k^{\ast} = \frac{\mu_{\mathrm{T}}k\sigma^{2}}{D_{\mathrm{T}}}    
\end{equation}
where $Pe$ is the P{\'e}clet number that is a measure of the self-propulsion in terms of the diffusion of motile particles, $A$ is an effective elastic dipole-dipole interaction parameter, and $k^{\ast}$ is the nondimensional steric spring constant.

Nondimensional forms of the dynamical equations Eq.~\ref{eq:translation}-\ref{eq:orientation} are discretized and numerically solved using the explicit half-order Euler-Maruyama method \cite{Allen2017}. We use a time step of $\Delta t = 10^{-4}$ for a total of $10^{5}-10^{6}$ timesteps corresponding to a total simulation time of $10-100$. Each particle was initialized with a random position and orientation in our simulation box of size $L_{x}=30\sigma$ and $10\sigma \leq L_{y} \leq 30\sigma$ with periodic boundary conditions in $x$ and confinement modeled by repulsive springs identical to those used for particle-particle steric repulsions, with a fixed spring constant, $k^{\ast}=10^4$, placed along the top and bottom walls. $A$ and $Pe$ are varied and analyzed in the main results section of the text.

\section{Discussion}
We have shown the typical collective behavior that emerges when active Brownian particles interact with each other as dipoles, using Brownian dynamics simulations. This  minimal model describes collective cell motility on elastic substrates where the cell-cell interaction is mediated by their mutual deformations of the passive substrate. 

The passive dipolar interactions lead to the end-to-end alignment of the particles into motile chains, which can be mutually aligned into polar bands and clusters because of their active motion. Polar chains that travel in opposite directions would be sorted into bands that get out of each others' way. These basic implications of our model, while specifically demonstrated here for elastic dipoles, belong to a broader class of active particles with dipolar interactions \cite{Kaiser2015, Liao2020}, and may therefore also be experimentally realized in active colloids endowed with permanent or induced dipole moments \cite{sakai2020active}.  We note that the symmetry of the elastic dipolar interactions is modified at higher Poisson's ratio \cite{Bischofs2003}, which is expected to result in structures such as active rings with rotational motion. This richer behavior with elastic interactions is a direct consequence of the tensorial, as opposed to vectorial nature of the elastic dipoles, in contrast with magnetic or electric dipoles, and will be the subject of future study.

We focused on the strong elastic interaction cases in the dilute regime, where the self-assembly and dynamics of single chains can be studied. Since the chains are stable in this regime, they resemble other active polymer systems \cite{Winkler2020}, that typically arise in gliding assays of biological filaments \cite{phillips} or with synthetic colloids \cite{Nishiguchi2018}. Polar bands are also seen at a higher density of active polymers \cite{Vliegenthart2020}. However, in our system where these chains are self-assembled by dipolar interactions, multiple chains can stick to each other at higher interaction strength, while they can also fall apart, when colliding at high motility. By showing how a pair of chains interact with each other, we show the stable higher order structures that form and contribute to the polar clusters seen at higher density. Although not investigated in detail here, it will also be interesting to explore  the bending dynamics of a single active polymer \cite{Eisenstecken2016, Gopinath2021} and characterize how the bending rigidity increases with dipolar interaction strength or decreases with particle motility. 

To conclude, we note that our cell mechanobiology-inspired model also realizes a new class of active matter with long-range dipolar interactions. The emergent self-organization behavior distinct from the two typically studied pathways to the clustering of active particles: motility-induced phase separation \cite{Cates2015}, and Vicsek-style models \cite{vicsek1995}. In the latter, particle alignment is imposed in an agent-based manner, whereas here alignment emerges as a natural consequence of physical interactions.

\section*{Conflict of Interest Statement}

The authors declare that the research was conducted in the absence of any commercial or financial relationships that could be construed as a potential conflict of interest.

\section*{Author Contributions}
SB and PN contributed equally to this work.
ArG, AjG and KD conceived the study and the model. SB and PN performed the numeric simulations and analyses. All authors wrote the manuscript.

\section*{Funding}
PN was supported by a graduate fellowship funding from the National Science Foundation: NSF-CREST: Center for Cellular and Biomolecular Machines (CCBM) at the University of California, Merced: NSF-HRD-1547848. ArG (Arvind Gopinath) acknowledges funding from the National Science Foundation via the CAREER award NSF-CBET-2047210. AjG (Ajay Gopinathan) also acknowledges partial support from the NSF Center for Engineering Mechanobiology grant CMMI-154857.

\section*{Acknowledgments}
PN and SB acknowledge computing time on the Multi-Environment Computer for Exploration and Discovery (MERCED) cluster at UC Merced (NSF-ACI-1429783) 
SB and PN acknowledge useful discussions with Farnaz Golnaraghi and Jimmy Gonzalez Nunez.



\bibliography{bibliography}

\end{document}